\documentstyle[aps, preprint]{revtex}
\begin{document}
\title{Relativity and EPR Entanglement: Comments}
\author{ S.C. Tiwari}
\address{ Institute of Natural Philosophy\\
1 Kusum Kutir, Mahamanapuri\\
Varanasi 221 005, India}
\maketitle        
\begin{abstract}
Recent experiment by Zhinden {\it et al} (Phys. Rev {\bf A} 63
02111, 2001) purports to test compatibility between relativity
and quantum mechanics in the classic EPR setting. We argue that
relativity has no role in the EPR argument based solely on 
non-relativistic quantum formalism. It is suggested that this 
interesting experiment may have significance to address fundamental
questions on quantum probability.
\end{abstract}


Authors of \cite{1} report first experimental realization of what
they call 'relativistic configurations' to test nonlocal quantum
correlations. In the vast literature on the subject, it is
routinely stated that quantum non locality does not violate relativity
as no information is communicated at superluminal speed. Since the
arguments are intuitive and philosophical, a scheme to test this
claim seems desirable. Zhinden {\it et al} \cite{1} use entangled
photon pair, set the detectors separated by more than 10 km in 
relative motion and measure the quantum correlations. They admit
possibility of questionable assumptions, hint at metaphysical 
considerations and invite constructive criticism. While these issues
are surely being debated, I ask a basic question: Should one invoke 
relativity in EPR setting ? If no, the motivation of this experiment
collapses addressing an ill-posed problem. In this brief comment we
analyze this question, and suggest alternative approach for 
interpreting/designing such experiments.

Original EPR incompleteness argument considers non-relativistic quantum
systems (A and B) which interact for a short time, and then evolve freely
obeying the Schroedinger equation \cite{2}. Canonically conjugate 
variables position and momentum are taken as observable. It is
shown that measurement on any one system allows prediction of the
observable(s) of the second without disturbing that system. It is
concluded that $\psi$ function does not provide a complete description
of the physical reality. Bohr argues that a quantum system exists
in all possible superposed states until the measurement is made, and
it is not justified to ascribe physical reality to any observable
till then. In the EPR scheme, Bohr's interpretation implying 
simultaneous collapse of the  $\psi$ function for the second system
has led to many speculations: non-local interaction between A and
B, superluminal communication between them, the role of mind etc.
For the distant correlations such that no signal with light speed can 
connect A and B, conflict with relativity/superluminal signals
become important issues. Thus this problem is tied with the Copenhagen
interpretation, and EPR did not raise relativistic questions in 
their paper. Since the quantum systems are non-relativistic, Lorentz
invariance or special theory of relativity is not applicable. As a 
matter of principle, there is no objection to assume instantaneous
action at a distance. It must be realized that intervening medium 
or mediating signal is not required in direct particle force in action
at a distance \cite{3}.

Let us now examine Bohm's version of EPR. It is intriguing that in all
discussions only entangled spin states are used while complete 
$\psi$- function should be product of the entangled spin states
and free particle plane wave functions. Plane wave has infinite
extension, during a measuring act in system A this can play the
role of non-local correlation with system B. Extending Bohm's version
to photon pairs leads to fundamental problems: 1) suitable Lorentz
covariant $\psi$-function for photon, 2) relativistic analogue
of Schroedinger equation for photon, and 3) consistent treatment of
Lorentz and gauge invariance. For a comprehensive review on them, 
see [4]. The best possible description of a photon in quantum optics
in terms of a creation operator acting on vacuum state also has a 
plane wave, therefore particle-like detection requiring photon localization
creates difficulties \cite{5}. However, for EPR setting the entangled 
states and  $\psi$-function for photons too are assumed to be of
non-relativistic form. Clearly this assumption is unjustified.

Though authors of \cite{1} are careful to point out that their
experiment is not  designed to test the model proposed in \cite{6},
they are inspired by the idea that not only space like separation but also
relative moving frames of reference are important. In contrast to \cite{6}
we note that space-time is absolute in relativity (space and time are relative),
and local realism is not the Einstein' assumption that 'nothing in physical
reality happens faster than light'. Quite often, non locality is defined
to mean the impossibility of a local hidden variable construction 
satisfying quantum mechanical statistical predictions. 
We follow original EPR arguments, the statement, '...(if) two systems 
no longer interact, no real change can take place in the second system 
in consequence of anything that may be done to the first system'
embodies the locality assumption in \cite{2}. The two systems A and 
B evolve independently: unitary time evolutions without any
mediating interaction. The act of measurement and correlated physical 
obseravables for two systems led EPR to conclude that quantum theory is
incomplete. Even in Bohm's hidden variable theory, 'non-contextual' is more
appropriate than 'local realism' \cite{7}, here non-contextual means that the
choice of a random variable for an observable is independent of 
simultaneous measurement of the other variables. Let us now analyse the 
role of relativity in the experimental setup of \cite{1}. Observers in
relative motion perform measurements on the space like separated quantum 
systems in this scheme.
Zhinden {\it et al} give a reasonable definition of the act of measurement:
irreversible record on detector- signifying the collapse of the wave 
function or quantum to classical transition.Detection is a classical 
event, and relative motion of detectors corresponds to the relativity 
of simultaneity in classical relativistic sense.Unless relativistic 
rendition of the wave function collapse is made,this experiment
does not say anything regarding quantum theory.Relativistic description
of EPR setting needs relativistic transformations for unitary evolutions
,definition of quantum clocks,and Lorentz transformed collapse process.
Thus the kind of experiment reported in \cite{1} cannot resolve 'tension'
between relativity and quantum theory, if any.

Lack of precision in the language and metaphysical elements in the
discussions on the foundation of quantum theory invite strong reactions 
\cite{8}. I admit that distilling physics from this paper was very difficult,
but I believe there is interesting physics in this experiment \cite{1}.
Following EPR's advice \cite{2}: 'The elements of the physical reality 
cannot be determined by a priori philosophical considerations, but must
be found by an appeal to results of experiments and measurement' we
suggest an alternative approach. Basic ingredients are 1) conservation
laws, and 2) coincidence statistics observed in the experiments originates
due to randomness in the pair creation at the source \cite{9}. Since
first ingredient is used in all the varied interpretations, we
elaborate on the second. In all the experiments, observational data
represent the ensemble of EPR pairs, but the theoretical discussions
consider a single pair and derive weird conclusions. Since the physical 
reality of quantum objects, {\it e.g.} photons in terms of concrete models
is not certain \cite{4}, there is always a room for ambiguous 
interpretations. Assuming that each photon in a pair has definite
physical properties independent of the intervening devices/measuring
instruments, conservation laws determine the correlated attributes.
Random generation of the pairs means, for example for the polarization
correlated photons, that photons in a pair lie on the diametrically opposite
points on the Poincare sphere and the states of the ensemble of the pairs
are distributed over the Poincare sphere. Using Pancharatnam's analysis phase
for a polarized state can be defined, and a probability distribution for 
phase is obtained \cite{10}. Though Suter \cite{10} used this classical 
analysis in the context of 'quantum time-translation machine', we propose 
that similar probability distributions for EPR pairs (at the source)
on the Poincare sphere (or sphere in k-space) can account for quantum
correlations. Weihs {\it et al} \cite{11} consider a 
loophole that only a small portion of the ensemble of pairs created
is detected; in their experiment detection efficiency was $5 \%$. Next
there is the question of time scales involved. Tentatively Zhinden {\it et al}
experiment appears to be suitable to throw light in this aspect. Time
intervals between signal and idler photons generation, see nice discussion
in section 22.4.7 in \cite{5}, and between successive pair creation should
show up in the measurements performed  with the detectors set in
relative motion. To conclude this approach need to be developed further in
conjunction with inputs from the recent experiments on EPR correlation
to demystify quantum theory.

\noindent {\bf Acknowledgment}

Library facility of the Banaras Hindu University, Varanasi is
acknowledged.

\end{document}